\documentstyle[11pt,newpasp,twoside,psfig]{article}
\index{instructions}
\index{guidelines}

\def\edcomment#1{\iffalse\marginpar{\raggedright\sl#1\/}\else\relax\fi}
\reversemarginpar

\begin{document}
\title{A new population of X-ray transients in the Galactic Centre}
 \author{Masaaki Sakano, Robert S. Warwick}
\affil{Dept. Phys. \& Astron., Univ. Leicester, Leicester LE1 7RH, UK
}
\author{Anne Decourchelle}
\affil{CEA/DSM/DAPNIA, C.E. Saclay, 91191 Gif-sur-Yvette Cedex, France}
%CEA/DSM/DAPNIA, Service d'Astrophysique
\author{Q. Daniel Wang}
\affil{Dept. Astron., Univ. Massachusetts, Amherst, MA 01003, USA}
% LGRT-B 619E, 710 North Pleasant Street, Amherst, MA 01003

\begin{abstract}
A comparison of the {\it XMM-Newton} and {\it Chandra}  
Galactic Centre (GC) Surveys has revealed two faint X-ray transients 
with contrasting properties. The X-ray spectrum of XMM J174544$-$2913.0 
shows a strong iron line with an equivalent width of $\sim$2~keV, 
whereas that of XMM J174457$-$2850.3
is characterised by a very hard continuum with
photon index $\sim$1.0.  The X-ray flux of both sources varied by more than 2
orders of magnitude over a period of months with a peak X-ray luminosity
of $5\times 10^{34}$erg~s$^{-1}$.  We discuss the nature of these %two
peculiar sources.
\end{abstract}

\section{Introduction}

The Galactic Centre (GC) probably harbours a great number of transient 
X-ray sources.  Past X-ray observations
have revealed that the majority of the transient sources with the luminosity of $L_{\rm X}
\geq 10^{35}$erg~s$^{-1}$ are low-mass X-ray binaries (LMXBs),
containing a neutron star or black hole (eg., Sakano et al. 2002).  Recent {\it Chandra} 
and {\it XMM-Newton} observations have lowered the detection threshold by
1--2 orders, thus providing access to potential new X-ray populations of sources
with luminosity in the range $L_{\rm X}=10^{32}$--$10^{34}$erg~s$^{-1}$.  
Here we report two relatively faint X-ray transients, which exhibit unusual properties.

\section{Results \& Discussion}

We have compared the {\it XMM-Newton}/EPIC and {\it Chandra}/ACIS Survey data 
obtained during the period between September 2000 and June 2002.  We detected several transients 
within the 0.3$\times$0.3~deg$^2$ field centred at ($l$,$b$)=($-0.\!^{\circ}1$,
$-0.\!^{\circ}2$).  Among them we detected XMM J174544$-$2913.0 and XMM
J174457$-$2850.3 at the respective J2000 positions of 
(RA, Dec) = ($17^{\rm h}~45^{\rm m}~44^{\rm s}\!.38$,
$-29^{\circ}~13'~0''\!.6$) and
($17^{\rm h}~44^{\rm m}~57^{\rm s}\!.56$,
$-28^{\circ}~50'~20''\!.7$)
coordinates with an error radius of 8$''$.

XMM J174544$-$2913.0 was detected in September 2000 ($L_{\rm
X}=5\times 10^{34}$erg~s$^{-1}$), but not in July or September 2001
with the lowest 3$\sigma$ upper limit for the 2--10 keV luminosity of
$3\times 10^{32}$erg~s$^{-1}$, assuming a distance of 8.0~kpc.  As for
XMM J174457$-$2850.3 we determined its 2--10 keV X-ray luminosity 
to be $1\times 10^{33}$, $5\times
10^{34}$, $1\times 10^{32}$erg~s$^{-1}$ in July 2001, September 2001
and May 2002 respectively.  Furthermore, in the July 2001 observation 
the source flux declined by a factor of two or more in 10~ks.

\begin{figure}[tbp]
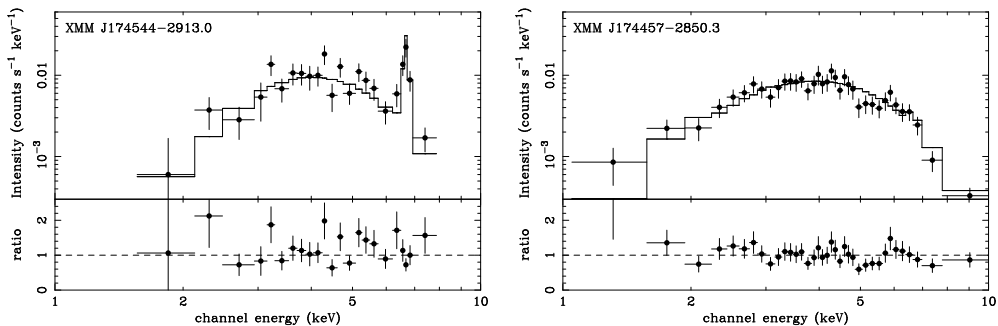

\begin{center}
{\small
\begin{minipage}[t]{0.48\textwidth}
  \begin{center}
  \mbox{\psfig{file=SakanoM1_1.ps,width=\textwidth,angle=270}}
%  \mbox{\psfig{file=j174544_rat.ps,width=\textwidth,angle=270}}
%\centerline{\psfig{file=tmp.eps,width=\textwidth}}
%  \caption[]{%{\it XMM} spectrum of
% XMM J174544$-$2913.0.
%\label{xmmfe}}
  \end{center}
\end{minipage}~~
\begin{minipage}[t]{0.48\textwidth}
  \begin{center}
  \mbox{\psfig{file=SakanoM1_2.ps,width=\textwidth,angle=270}}
%  \mbox{\psfig{file=j174457_rat.ps,width=\textwidth,angle=270}}
%  \caption[]{%{\it XMM} spectrum of
% XMM J174457$-$2850.3.
%\label{xmmtr}}
  \end{center}
\end{minipage}
  \caption[]{%{\it XMM} spectrum of
% XMM J174544$-$2913.0 (Left) and J174457$-$2850.3 (Right).
 {\it XMM-Newton} MOS1 spectra of the two transient sources.
\label{fig:spec}}
}
\end{center}
\end{figure}

Fig.~\ref{fig:spec} show the {\it XMM} spectra of the two sources when
they were in the high state.  XMM J174544$-$2913.0 was found to have an
extremely strong iron line with a centre energy of 6.68$\pm$0.02 keV and
an equivalent width (EW) of 2.4$^{+0.4}_{-0.5}$ keV, whereas XMM
J174457$-$2850.3 exhibits a very hard continuum  with photon index of
0.98$^{+0.33}_{-0.25}$ with a weak 6.7-keV line (EW=180$\pm$140 eV).
Both the spectra are absorbed by a large column density: 12.4$\pm$1.8
and 5.9$\pm$1.1 $\times 10^{22}$ H~cm$^{-2}$, respectively.  XMM
J174457$-$2850.3 showed marginal evidence for softening of the spectrum
from the high state ($\Gamma\sim 1.0$) to low state ($\sim 1.9$).

The strong iron line and transient nature of XMM J174544$-$2913.0 is
quite similar to AX J1842.8$-$0423 (Terada et al. 1999).  Thus, as
suggested by Terada et al., it is likely to be a magnetised cataclysmic
variable (CV) viewed from a pole-on inclination, which causes an apparently
strong line at 6.7 keV from helium-like iron.  However the large
luminosity of over $10^{34}$erg~s$^{-1}$ is quite unusual for CVs and some
additional component, for example a jet, may contribute to the observed emission.

The nearly featureless and flat spectrum of XMM J174457$-$2850.3, as
well as the existence of diffuse emission around the source, suggests
that it may be a neutron star or black hole binary.  The weak but
significant iron line and the flat index point to this being
a high mass X-ray binary (HMXB). However, both the quiescent luminosity 
of $1\times 10^{32}$erg~s$^{-1}$ and  the peak observed luminosity 
of $4\times 10^{34}$erg~s$^{-1}$ are unusually low, suggesting the
possibility of a wide eccentric orbit characteristic
of many Be star X-ray binary systems.

\end{document}